\documentclass[11pt,final,titlepage,a4paper]{article}
\usepackage{latexsym}
\begin{document}

\begin{titlepage}
\title{Oligopeptides' frequencies in the classification of proteins' primary 
structures.} 
\author{ Paolo Sirabella 
\textsuperscript{\dag \ddag}, Alessandro Giuliani \textsuperscript {\S}
, Alfredo Colosimo \textsuperscript {\dag} \\
\footnotesize  \dag \ University of Rome "La Sapienza" - Dept. of Biochemical Sciences \\
\footnotesize P.le A. Moro, 5 - 00185 Roma - Italy \\
\\
\footnotesize \S \ Istituto Superiore di Sanit\'a - TCE Lab. \\
\footnotesize V.le Regina Elena,199 - 00161 Roma -Italy \\
\\
\scriptsize \ddag \ To whom correspondence should be addressed. \\
\scriptsize Email: p.sirabella\char 64 caspur.it - Fax: ++39 06 49910957\\
\\
\vspace{20 mm}
\\
\footnotesize \textbf{Running title:} Oligopeptides in the classification of proteins.\\
\\
\footnotesize \textbf{Keywords:} Proteins' classification, SOM algorithm, Aminoacids' coding \\
\\
\footnotesize \textbf{Abbreviations used in the text:} \\
\footnotesize AAs: aminoacids; PCA: Principal Component Analysis; \\ 
\footnotesize SOM: Self Organizing Maps; SMH: Set Mean Homology; MMH: Map Mean Homology.
\date{\today}
}

\maketitle{}

\begin{abstract}

This paper reports about an approach to the classification of 
proteins' primary structures taking advantage of the Self Organizing 
Maps  algorithm and of a numerical coding of the
aminoacids  based upon their physico-chemical properties.

Hydrophobicity, volume, surface area, hydrophilicity, bulkiness, 
refractivity and polarity were subjected to a Principal Component 
Analysis and the first two principal components, explaining 84.8 \% of 
the total observed variability, were used to cluster the aminoacids 
into $4$ or $5$ classes through a {\em k-means} algorithm.  This leads 
to an economical representation of the primary structures which, in 
the construction of the input vectors for the Self Organizing Maps 
algorithm, allows the consideration of up to tri- and tetrapeptides' 
frequency matrices with minimal computational overload.

In comparison with previously explored conditions, namely symbolic 
coding of aminoacids and dipeptides frequencies, no significant 
improvement was observed in the classification of 69 cytochromes of 
the \emph{c} type, characterized by a high degree of structural and 
functional similarity, while a substantial improvement occurred in the 
case of a data set including quite heterogeneous primary structures.

\end{abstract}
\end{titlepage}

\section{Introduction}

Coding the primary structure of proteins by lists of numbers related to
the physico-chemical properties of the aminoacids (AAs)  in the polypeptide
chains should provide substantial help in the study of the correlations
between primary and tridimensional structures (Eisenhaber et al. 
1995;   Rost and Sander 1993; Reyes et al. 1994),  and
hopefully shade some light on the intricacies of the rules governing
proteins' folding (Fedorov and Baldwin 1997).

Although the issue is in the literature since a long time 
(Argos 1987; Schneider and Wrede 1993),  
the vast majority  of the software tools devoted  to the analysis of the primary
structure  (Thompson et al. 1994; Wishart et al. 1994)  utilize the
symbolic coding  of AAs, the main reason  being the successful drawing
of phylogenetic  trees  on the basis of homologous proteins of different
species after proper alignment (Page 1996).

Numerical coding of aminoacidic residues on solid physico-chemical and
statistical grounds, however, allows to take advantage of a manifold of
numerical multivariate data-analysis techniques and, in particular, to
fully exploit the euristic power of automatic
classification based upon Self Organizing Maps ({\em SOMs}), introduced
by Kohonen  several years ago (Kohonen 1984) as a general purpose tool
for classifying the elements  of a multivariate set.
The only strict requirement  of their unsupervised learning mechanism,
i.e. the same number of variables  for each element of the set, can be
easily met even if the primary structures to be classified are of 
different length. To any protein, in fact, can be associated  a
frequency matrix of  $n^{d}$ elements, where  each element  is the
number of occurrences of each of the possible oligopeptides of length
$d$ within the primary structure ($n=20$ in the case of the natural
AAs).
On the basis of this approach, 
assuming a different symbol for each of the $20$
natural aminoacids and $d=2$, i.e. generating  frequency matrices  of
$20^2$  elements, it was possible to carry out both  fine classifications
within  sets  of structurally  similar proteins (Ferr\'an and Ferrara 
1991,1992),  
and  coarser  classifications over much larger sets (Ferr\'an et al. 
1992).

If, on one hand, increasing the length $d$ of the oligopeptide accounts
with higher and higher precision for the fine details of  each
individual primary structure, the exponential increase in the number of
the possible $d$-plets in the $n^{d}$ matrix poses some practical
and theoretical limitations. The former ones obviously refer to the
computational load, while the latter are related to the linearly
decreasing number of oligopeptides ($N-d+1$) with wich a sequence of
length $N$ may contribute to the non-zero, i.e. significant, elements of
the frequency matrix.

In this paper two exemplary cases of proteins' primary structure
classification will be described in which an appropriate balance between
the $n$ and $d$ values in the frequency matrices  feeding  the
{\em SOM} algorithm allows to: \textbf{i)} use oligopeptides longer
than dipeptides as descriptors  of the primary structures, and
\textbf{ii)} minimize the ensuing computational load by  lowering the
size of $n$  with no (or minimal) loss of the statistically significant
information, through the combined  use of principal component and
cluster analysis techniques.

\section{Methods}

\subsection{Self Organizing Maps  (SOM)}

The {\em SOM} algorithm, proposed by Teuvo Kohonen in the first 80s 
(Kohonen 1984), is a fully automatic algorithm that drastically 
reduces the dimensionality of a highly multivariate data set still 
preserving the mutual correlations between its elements.  The most 
recent implementation of such algorithm ({\em SOMPAK 3.1}, free 
software available, together with a rich bibliography, at the Web site 
\texttt{http://www.cis.hut.fi/nnrc/}) has been used throughout the 
present paper.

In our case the input of the algorithm is a set of numerical vectors 
obtained by an appropriate recoding of the primary structures of 
proteins, and the output is a bidimensional map where the mutual 
locations of the primary structures reflect their intrinsic 
similarities.  An extensive and clear description of the algorithm's 
working machinery is available in the literature (Kohonen 1995), where 
an estimate of the distorsion introduced in the original structure of 
the data set by reducing their dimensionality is given in the form of 
a {\em stress factor}.  As a more specific index of the goodness of 
the classification obtained in the case of proteins, the {\em Map Mean 
Homology} ($MMH$) index (see below) has been used throughout this 
paper.

\subsection{Calculation of the $MMH$ (Map Mean Homology) index}
\par
To evaluate the goodness of the clustering provided by the {\em SOM}
algorithm, the {\em Map Mean Homology} ($MMH$) index has been used,
along the same line followed by Ferr\'an and Ferrara (1991).
Such index can be defined as

\begin{equation}
MMH =
\frac{\displaystyle\sum_{i=1}^n {QR_{i}}^{Clusters}}{n}
\label{MMH}
\end{equation}

\noindent
i.e. the average of the Quality Ratio values (${QR_{i}}^{Clusters}$ values)
associated to the $n$ clusters present on the map. A cluster is defined
by the presence in a cell of at least two elements, and is extended to
its first neighbours, counted only once. Thus, the ${QR_{i}}^{Clusters}$ for
the $i^{th}$ cluster is defined as

\begin{equation}
{QR_{i}}^{Clusters} =
\frac{\displaystyle\sum_{j=1}^m W_{j}{QR_{i,j}}^{Couples}}
{\displaystyle\sum_{j=1}^m W_{j}}
\label{QRCL}
\end{equation}

\noindent
where $j$ runs over the $m$ couples associated to the $i^{th}$ cluster
and to its first neighbours, weighted by $W_{j}$ values of $1$ and $0.5$ 
in the former and latter case, respectively.

\subsection{Principal Component Analysis of AAs' physico-chemical 
properties.}
\par
The Principal Component Analysis (PCA), introduced by Pearson in 1901,
is a method of decomposing a correlation or covariance matrix  in order
to find the best association of points in space (Jolliffe 1986).

The first goal of the principal components is to summarize a
multivariate data set  as accurately as possible using  fewer
uncorrelated variables. This can be achieved since the principal
components are orthogonal to each other, thus removing any  redundancy
in the available information. The relation between the original
variables and the principal components is expressed in terms of
\emph{component loadings}, i.e. the correlation coefficients of the
original variables with the new ones (principal components).

In this paper PCA has been carried out over seven physico-chemical 
properties of the $20$ natural AAs, namely hydrophobicity, volume, 
surface area, hydrophilicity, bulkiness, refractivity index and 
polarity which, according to Schneider and Wrede (1993), are relevant 
in the identification of specific patterns along proteins' sequences.  
Among these properties, hydrophobicity has been recently confirmed as 
by far the most important one in protein folding (Weiss and 
Herzel 1998).  In Table \ref{loadings} our PCA results are reported 
in terms of the components' loadings and of the percent variability 
explained by each component.  The first and second  
components (PC1, PC2) explain $84.8\% $ of the
 total variability and hence have been  considered as 
 reliable and non redundant  representatives of the whole set of
 properties.

\subsection {k-means clusterization of AAs.}
 The {\em k-means} algorithm  is  a semi-automatic procedure to
identify classes within a given set of elements described by one or many
variables (Everitt 1980).
Clusters emerge here from the structural characteristics of the data
set, by maximizing the interclass variance and minimizing the intraclass
variance. For  $n$ units described by $m$ variables, the procedure can
be schematized as follows:

\begin{enumerate}
\item  a non-trivial number of classes, $k$, is defined, being 
$1<k<n$;
\item  $k$ aggregation points in an $m$-dimensional space are
       arbitrarily chosen;
\item  each of the $n$ units is assigned to the nearest aggregation point;
\item  a new set of aggregation points are reckoned as barycentres
       of the classes defined in the previous step;
\item  go back to the $3^{rd}$ step until no further change occurs in the
       classes' composition.
\end{enumerate}

The external factor which makes the procedure non fully automatic, is
the \textit{a priori} definition of $k$. 

In the present case, the
algorithm has been used  to group into $k$ classes the $20$ AAs
on the basis of their hydrophobicity ($m=1$), as well as the values of
the first and second principal components  ($m=2$) extracted from their
main physico-chemical properties.

The relative optimality of the $k$ value can be chosen by means of the 
relation between the fraction of explained variability ($EV$)  
relative to the classification, and the value of $k$: reaching a plateau of 
$k$  can be considered the result of a structurally optimal classification 
(see also the legend to Table \ref{clusters}).

\section{Results}

\subsection{Data sets used in this paper}
The leading criterium in the choice of the two data sets used in this
work reflects the aim to test the performance of a numerical coding  of
the AA and of a variable length of the oligopeptides' describing the primary 
structures  under two different conditions, namely a  high and a low 
value of a global  similarity index (see below).

For Data Set I, shown in Table \ref{dataset1}, were chosen $69$ 
cytochromes of the \emph{c} type, which are known to share a high 
level of both structural and functional similarity.  To impose some 
rational constraint in the choice of Data Set II, where a high 
similarity in the primary structures was not a prerequisite, our 
attention focussed over a group of proteins in which, as shown by 
Alexandrov and Fisher (1996), a significant similarity in the 
tridimensional arrangements was unparalleled by any homology in the 
primary structures.  The elements of Data Set II are listed in Table 
\ref{dataset2}.

It is worth stressing that the two data  sets should be considered from two 
complementary viewpoints:

\textbf{i)} since the differences between the elements in Data Set I 
consist in a number of gaps/point-mutations over essentially the same 
basic primary structure, any source of variability (information) 
related to structural and/or functional features, is expected to be 
minimal.  Under these conditions even the simplest symbolic coding 
 blind to physico-chemical features can be appropriate;

\textbf{ii)} the high heterogeneity within the elements of Data Set 
II, related to their quite different length, composition, function and 
primary structure, should be in favour of any classification task 
based on a numerical coding of the sequences.  This introduces, 
however, new problems about choosing the optimal physico-chemical 
descriptors of the AAs, or about how to group them into clusters, on which 
heavily depends the classification's goodness. 

A quantitative estimate of the {\em Set
Mean Homology} ($SMH$) among the $n$  elements (in couples) within a
set is given by

\begin{equation}
SMH =
\frac{\displaystyle\sum_{i=1}^{n-1}\displaystyle\sum_{j=i+1}^{n}QR_{ij}}
{n(n-1)/2} \label{SMH}
\end{equation}

\noindent where the \emph {$QR_{ij}$} are the $Quality Ratio$ values, 
i.e.  correspond to the elements of a triangular matrix generated as 
an intermediate result by the {\em PILEUP} program in the {\em GCG} 
suite of programs for the analysis of biosequences (Doelz 1994).  More 
precisely, each $QR_{ij}$ is given by an estimate of the goodness of 
the alignment between the $i,j$ elements in the data set as provided 
by : \textbf{i)} the Needleman-Wunsch algorithm (1969), and 
\textbf{ii)} a substitution matrix of the \emph{BLOSUM} type (Henikoff 
and Henikoff 1992), normalized by the number of residues of the 
shortest sequence between $i$ and $j$.  Notice that the procedure used 
in reckoning $QR_{ij}$ refers to a symbolic coding of the natural AAs, 
i.  e.  matches the condition used as a reference (black bars) in 
Figure 2.  However, high-quality classifications of primary structures 
can also be obtained upon clustering the AAs into $4$ or $5$ groups 
through a k-means algorithm, after an appropriate numerical coding 
provided by PCA.

\subsection{Classification of the data sets` elements.}
Figure 1 shows the map generated by the {\em SOM} algorithm in the 
case of Data Set I. This data set, due to the high level of similarity 
between the primary structures, constitutes a significant benchmark to 
test the fine discrimination power of the algorithm.  A very similar 
data set has been successfully analyzed by Ferr\'an and Ferrara (1992) 
using a symbolic coding of the 20 natural AAs and dipeptide 
frequencies, i.e. a vector of ($20^{2}$) components for each primary 
structure.  At difference with these authors, we used a numeric coding 
for the AAs  in the aim to: \textbf{i)} exploit the 
physico-chemical information characterizing each single residue; 
\textbf{ii)} increase the length of the oligopeptides; \textbf{iii)} 
minimize the computational burden by reducing the number of classes in 
which the residues can be clustered.  The main goal was to provide a 
more direct correlation between primary and tertiary structures' 
similarities.

A glance at Figure 1 indicates that even using vectors of $5^{3}$ 
components, corresponding to tripeptides' frequencies and to clustering 
the AAs into five groups, in the description of the primary structures, the 
phylogenetic relationships within cytochromes are very well preserved.

A quantitative estimate of the classification goodness obtained by the
{\em SOM} algorithm is provided in Figure 2 in terms of the Map Mean Homology 
($MMH$, see methods) score for both Data Sets I and II. In each panel 
of Figure 2 is also indicated (dotted line) the Set Mean Homology 
($SMH$, see Methods), i.e. an estimate of the overall similarity 
between all the couples of elements in the set.  Under all conditions 
the bars' heigth exceeds the dotted line of an amount indicating the 
performance of the classifier algorithm.  The bars in Figure 2 
represent the values of the $MMH$ for various combinations of: 
\textbf{i)} the coding criteria for the AAs; \textbf{ii)} the number of 
groups in which the AAs are clusterized; \textbf{iii)} the length of 
the oligopeptides whose frequencies constitute the vectors associated 
to each sequence.

The most interesting result provided by our analysis is the striking 
difference in the efficiency of the adopted coding scheme for the AAs, 
between the two data sets.  Taking as a reference the previously used 
symbolic coding coupled to dipeptide frequencies (black bars in Figure 
2), substantially identical results have been obtained under all 
conditions when the data set included elements of high $SMH$ (Figure 
2A).  Upon collapsing the latter constraint, however, a numerical 
coding based upon a PCA of their main physico-chemical properties 
(Table \ref{loadings}), and the ensuing techniques of clustering the 
AAs (Table \ref{clusters}) into $4$ or $5$ groups, provided a worse 
performance and a better one in the case of, respectively, dipeptides 
and tripeptides frequencies (Figure 2B).

To rationalize these results two basic points should be taken into 
account: first of all, it is quite obvious that, in very general 
terms, the ability of the {\em SOM} algorithm in finding peaks of 
similarity over a background of globally low similarity in the map is 
exalted.  Such an effect is independent from the coding criteria of 
the residues and only deals with the specific features of the elements 
to be classified.  It can be described by the expression:

\begin{equation}
\frac{<MMH> - SMH}{SMH}
\label{IMPROVM}
\end{equation}

\noindent
which, for the data shown in Figure 2A and B, gives the average values
of $0.19 \pm 0.03$  and $1.83 \pm 0.97$,  respectively.

Second, the much higher relative variance associated to the results in 
Figure 2B clearly indicates that the role of the coding criteria, 
namely \textbf{i)} oligopeptide length, and \textbf{ii)} optimized 
(through PCA) physico-chemical information, is only 
emerging in the case of Data Set II.

Finally, special consideration deserves the  difference observed
between the two data sets  when the classification occurs after a
\emph{random clustering} of the AAs in $4$, $5$ or $10$ groups (white
columns in Figure 2). Such a condition has been included in our
analysis to clarify the relative importance of the symbolic coding of
AAs (see Discussion).

\section{Discussion}

In classifying proteins of  different length on the basis of their
polypeptide sequences a crucial problem consists in the appropriate
coding of  the AAs, since  the appropriate statistical and
connectionist procedures usually require as an  input  numerical vectors
of identical dimension.  To overcome the problem  a "units-variables"
matrix may be worked out, where the rows are  associated to the proteins
and the columns contain, for example,  the  relative frequencies of the
$20$ natural AAs, or of dipeptides, tripeptides,  etc.,  thus providing
a more and more accurate (although longer) global  description of the
primary structures.
In particular, such an approach has been applied in the use of a neural
classifying algorithm, the \emph{SOM} (see Materials),  endowed with an
automatic features' extraction ability in the absence of any indipendent
information (unsupervised learning), with a minimum number of adjustable
parameters.

In this paper we showed that a synergic use of multivariate 
statistical techniques and of the \emph{SOM} algorithm is very 
effective, mainly in the case of heterogeneous data sets, given an 
appropriate choice of the coding criteria for the AAs and of the 
length of the oligopeptides used to represent the primary structures.  
This clearly appears from the comparison of the upper and lower panels 
in Figure 2, referring to data sets of high and low mean homology, 
respectively.  Under the former condition, as indicated by 
the high value of the $SMH$, all the explored criteria for primary 
structures's coding look almost equivalent.  The improvement obtained 
with reference to the more traditional symbolic representation of AAs 
and dipeptides' frequencies is  evident in the lower panel, 
where the data set includes elements of much lower $SMH$.

This poses the question whether a further improvement could be 
obtained by further increasing the oligopeptides length, $d$.  For both 
data sets used in this work this was actually not the case (not 
reported).  The main reason is related to the exponential increase, 
with increasing $d$, of the size of the frequency matrices, coupled to 
a linear decrease in the number of oligopeptides associated to each 
primary structure of length $N$ described over an alphabet of $n$ 
different symbols ($n=20$ for an unreduced symbolic representation of 
the $20$ natural AAs).  In other words, the ratio

\begin{equation}
\frac{N-d+1}{n^d}
\label{LIMIT}
\end{equation}

\noindent which represents the fraction of the non-zero elements in 
the frequency matrix for each polypeptide sequence, tends very rapidly 
to zero with increasing $d$.  Thus, the sparsity of the cumulative 
matrix related the whole data set, obtained from the element by 
element sum of the individual frequency matrices, should be considered 
as the main factor affecting the efficiency of the \emph{SOM} 
classifier.  Reducing to more favourable values expression \ref{LIMIT} 
by reducing $n$, i.e.  clustering the AAs residues into relatively 
homogeneous groups, needs the adoption of a numerical coding for the 
residues, on the basis of their hydrophobicity (Cid 1982; Reyes 1993) 
or, even better, of the principal components extracted from a bulk of 
physico-chemical properties.  The optimal number of such groups can be 
defined, in any case, through the Explained Variability index (see the 
legend to Table \ref{clusters}).  A complementary approach obviously 
consists in an appropriate filtering of the sparse matrices.

A possible objection to the above sketched strategy could invoke the 
observed insensitivity to the various coding schemes in the 
classification of the primary structures included into Data Set I. This focusses 
our attention on the peculiar features of the elements of this data 
set, namely on their structural (at the tridimensional level) and 
functional homogeneity, which seems to pose an intrinsic limit to any 
substantial improvement in the classification, even by increasing the 
oligopeptides' length.  It was not possible in fact, under the 
explored conditions, to outperform the traditional symbolic coding of 
the residues coupled to dipeptide frequency matrices.  A crucial 
observation in that respect, however, is that even after random 
grouping the residues into $4$ or $5$ classes the quality of the 
classification, as judged by the $MMH$ index, was not decreased. This 
points to the conclusion that even a relatively poor symbolic coding 
is able to capture the only relevant source of information in this 
peculiar data set, which could be associated to a variability of 
\textit{syntactic} type, i.e.  related to local differences between 
the elements of the set (relatively) independent from their 
macroscopic function, since all of them share a common structural and 
functional backbone (Yockey 1977).  In the absence of such common 
backbone, like in the case of Data Set II, where the substantial 
differences between the primary structures, give rise to a more 
\textit{semantic} (i.  e.  related to macroscopic functional 
differences) variability, the numeric coding of AAs should be 
preferred to the symbolic one.  It makes easier, in fact, by getting 
rid of the redundant information, to increase the length of the 
oligopeptides describing the primary structures, and hence a more 
accurate description of their global architecture, with substantial 
savings in terms of computational requirements.

 Up to what extent it is really worth to extend such length remains an open
 question. On the basis of a symbolic coding of the AAs, Strait and Dewey
 argued recently (1996) that the Conditional Information 
 Entropy  ($I_{k}$)  of k-tuples of AAs, used to estimate the Information Entropy ($I$)
 of proteins' primary structures through the expression
  
  \begin{equation}
    	I = \lim_{k\to \infty}I_{k}
    	\label{SHANN}
   \end{equation}
  
  already reaches a limiting value for $k$ equal to four.  Among other things,
  these authors are also able to work out a figure for the fraction of 
  the
  Information Entropy related to the tridimensional structure.  Thus, it
  seem of great interest to check their theoretical conclusions against the
  results of  an empirical approach based on the performance of \emph{SOM}
  classifiers and a physico-chemical coding of the AAs.
  
  \vspace{10 mm}
  
 \section{Acknowledgements}

Prof.  Ernesto Capanna and dr.  Stefano Pascarella, both from the 
University of Rome "La Sapienza", are gratefully acknowledged for many 
useful discussions.  This work has been partly supported by funds from the 
italian M.U.R.S.T. (40\% and 60\%).

\newpage

\textbf{References}

\begin{footnotesize}
\begin{list}{}{}
\item
Alexandrov NN, Fischer D (1996) Analysis of topological and nontopological structural 
similiarities in the PDB: new examples with old structures. Proteins Struct. Funct. Genet. 25:354-365
\item
Argos P (1987) A sensitive procedure to compare Amino Acid Sequences. J.Mol.Biol. 193:385-396.
\item
Bryant SH, Altschul SF (1995) Statistic of sequence-structure threading. Curr. Opin. Struct. Biol. 5:236-244
\item
Cid H et al. (1982) Prediction of secondary structure of proteins by means 
of hydrophobicity profiles. FEBS Lett. 150:247-254
\item
Doelz R (1994) Computer analysis of sequence data, part I (In) Methods 
in molecular biology, Vol. 24, Griffin AM, Griffin HG Eds. Humana Press, Totowa pp:9-171
\item
Eisenhaber F et al. (1995) Protein structure prediction: recognition of 
primary, secondary, and tertiary  structural features from amino acid sequence. Crit. Rev. Biochem. Mol. Bio. 30:1-94
\item
Everitt B (1980) Cluster Analysis. Halsted press, New York
\item
Fedorov AN, Baldwin TO (1997) Cotranslational protein folding. J. Biol. 
Chem. 272:32715-32718
\item
Ferr\'an EA, Ferrara P (1991) Topological Maps of Protein Sequences. 
Biol.Cybern. 65:451-458.
\item
Ferr\'an EA, Ferrara P (1992) Clustering proteins into families using artificial 
neural networks. Comput. Appl. BioSci. 8:39-44
\item
Ferr\'an EA et al. (1992) Large scale application of neural network to protein 
classification. Art. Neur. Net., Vol. II, North-Holland, pp:1521-1524
\item
Henikoff S, Henikoff JG (1992) Amino acid substitution matrices from protein blocks.
Proc. Natl. Acad. Sci. U.S.A. 89:10915-10919
\item
Jolliffe IT (1986) Principal Components Analysis. Springer-Verlag, New York, USA
\item
Kohonen T (1984) Self-organization and associative memory. Springer-Verlag, 
Berlin 
\item
Kohonen T (1995) Self-organizing maps. Springer-Verlag, Heidelberg 
\item
Needlemann SB, Wunsch CD (1969) A general method to the search for 
similarities in the amino acid sequences of two proteins. J. Mol. Biol. 
48:443-453
\item
Page RDM (1996) TreeView: an application to display phylogenetic trees 
on personal computers. Comput. Appl. BioSci. 12:357-358
\item
Reyes VE et al. (1994) Prediction of structural helices with the 
strip-of-helix algorithm. J. Biol. Chem. 264:12854-12858
\item
Rost B, Sander C (1993) Improved prediction of protein secondary 
structure by use of sequences profiles and neural networks. 
Proc. Natl. Acad. Sci. USA. 90:7558-7562
\item
Schneider G, Wrede P (1993) Development of artificial neural filters for 
pattern recognition in protein sequences. 
J. Mol. Evol. 36:586-595
\item
Strait BJ, Dewey TG (1996) The Shannon information entropy of protein 
sequences. Biophys. J. 71:148-155
\item
Thompson JD et al. (1994) CLUSTAL W: improving the sensitivity of progressive multiple sequence 
alignment through sequence weighting, position-specific gap penalties and weight matrix choice. 
Nucleic Acid. Res. 22:4673-4680
\item
Weiss O, Herzel H (1998) Correlations in protein sequences and 
property codes. J. Theor. Biol. 190:341-353
\item
Wishart DS et al. (1994) SEQSEE: a comprehensive  program suite for protein sequence analysis. 
Comput. Appl. BioSci. 10:121-132
\item
Yockey HP (1977) On the information content of cytochromes c. J. Theor. 
Biol. 67:347-376

\end{list}
\end{footnotesize}


\newpage

{\large \textbf{Legends to  tables and figures}}

\vspace{15 mm}

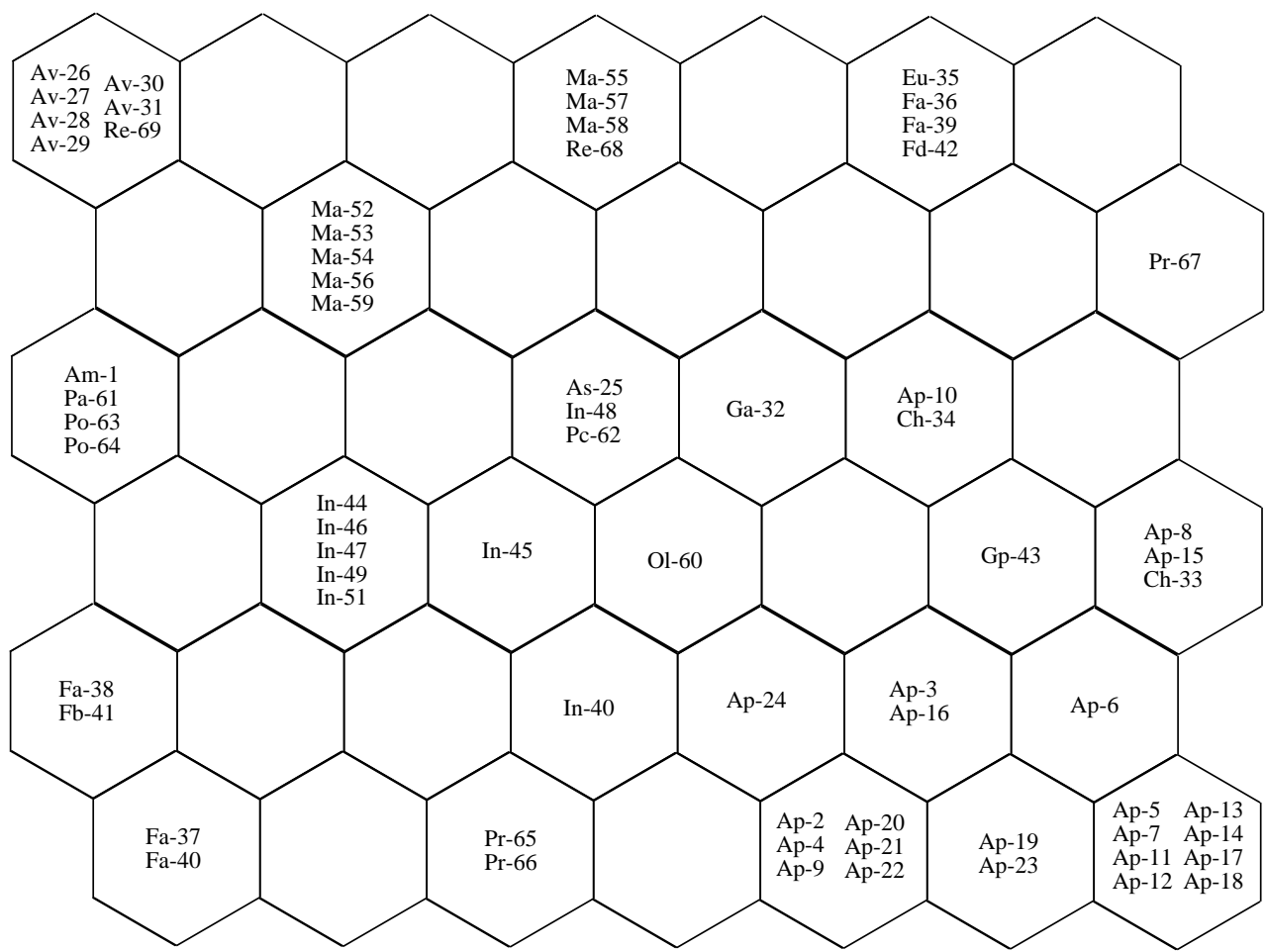
\textbf{ Figure 1: Classification of cytochromes of the \emph{c}
type by a {\em SOM} algorithm.}
\par
The map is a graphical rearrangement of the output provided by the \emph{SOMPAK 1.3} program
(see the text) on the cytochromes listed in Table \ref{dataset1}.
The input vectors, containing $5^3 = 125$ components, have been constructed:
\textbf{i)} grouping the AAs into $5$ classes by a \emph{k-means} algorithm on the basis of the
first and second principal components extracted from 7 physico-chemical 
properties (see the text), and \textbf{ii)} using the tripeptides` frequency
matrices.

The hexagonal lattice of the map and its overall size ($6 x 7$ cells) 
are a compromise between the conditions used by Ferr\'an and Ferrara 
(1992) and the Kohonen's suggestion to use different sizes 
for the map's axes.
The working parameters of the \emph{SOMPAK} program are the following:

{\small Lattice topology: hexagonal;\ Neighborood: bubble

\underline{First ordering phase:}

learning rate = $0.05$, $1000$ epochs, starting
radius $7$

\underline{Fine tuning phase:}

learning rate = $0.02$, $10000$ epochs, starting
radius $2$.
}

The maps refers to  the best results obtained, in terms of the internal distortion 
parameter, over $40$ different choices of the random
initial conditions  (see the Kohonen refs. for details)

\vspace{15 mm}

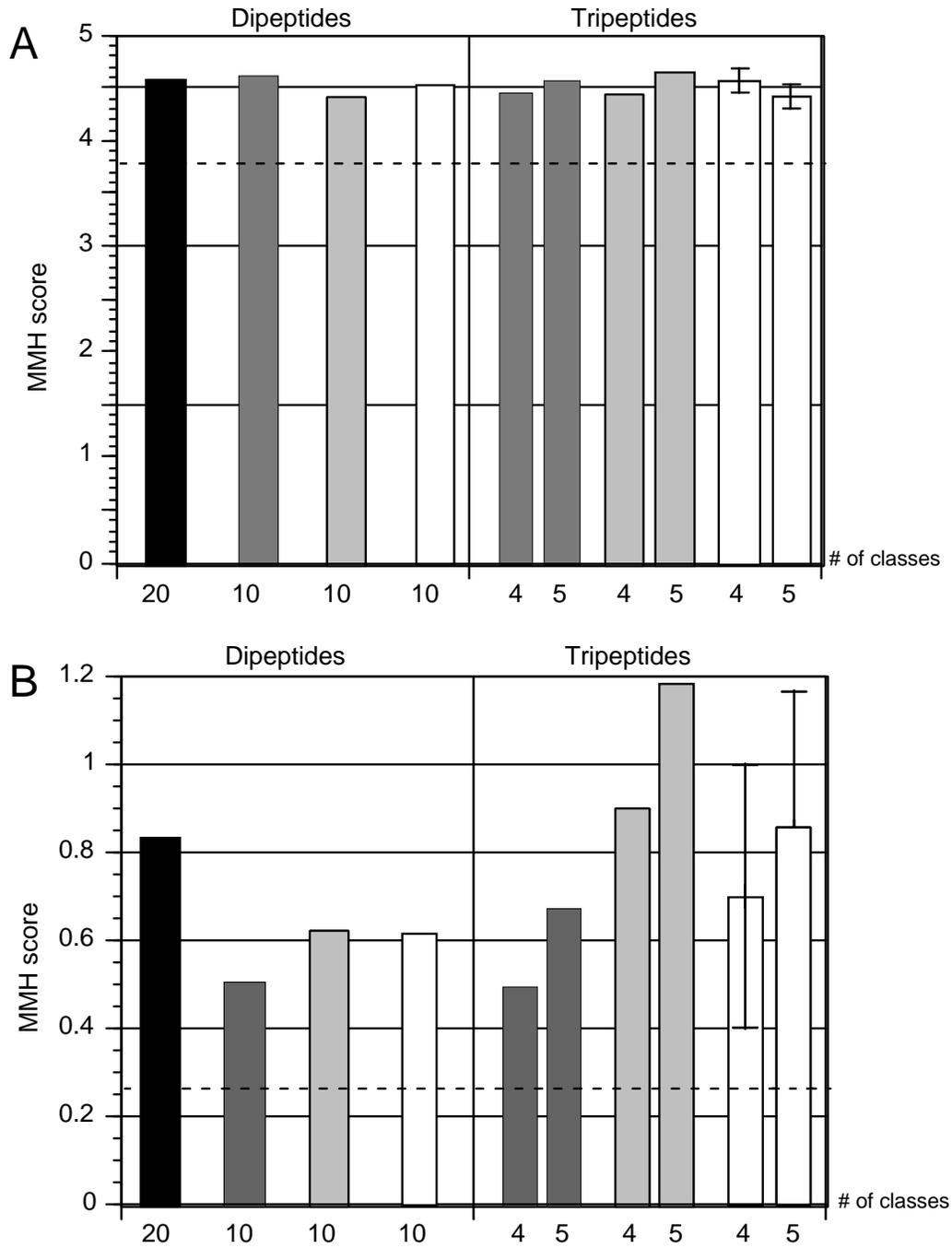
\textbf{Figure 2:  Performance of the \emph{SOM} algorithm for proteins'
classification under various conditions.}

\par
Panels A and B refer to the proteins in Data Sets I and II (listed in
Table \ref{dataset1} and Table \ref{dataset2}) and the histograms 
represent the $MMH$ (\emph{Map Mean
Homology}) score as defined in the text. The working parameters of the 
\emph{SOMPAK} program are the same listed in Figure 1 except that, in the case of
Data Set II, the dimension of the maps was $5 x 4$ due to the lower
number of elements.

The black, and white, bars refer to the unclustered natural, and 
randomly clustered AAs, respectively. The darker and lighter grey bars 
refer to clustering by hydrophobicity and, respectively, the PC1 + PC2 
extracted from physico-chemical properties (see the text). In the 
case of random clustering  each bar is the average of ten 
randomizations and the error bars indicate one standard deviation.

 \vspace{15 mm}
 
  \textbf{Table \ref{loadings}: PCA on seven physico-chemical 
 properties of the natural AAs.}
 
The table shows the correlations (loadings) between seven 
physico-chemical properties taken from Schenider and Wrede (1993) and 
the principal components extracted from them.  The first row reports 
the percent of the total variability $(EV\%)$ of the whole set of 
properties explained by each component.

\vspace{15 mm}

\textbf{Table \ref{dataset1}: Cytochromes of the \emph{c}  type used as Data Set I.}
 
Column $1$ is a numeric identifier for the corresponding entrance, 
without the {\em cytc} prefix, in the \emph{SwissProt} data-base 
(column 2).  Columns $3$ and $4$ refer, respectively, to the 
biological origin and the number of residues of each protein.  The 
used family abbreviations are the following: Amphibia (Am), Angiosperm 
(Ap), Asteroidea (As), Birds (Av), Gastropoda (Ga), Chlophyceae (Ch), 
Euglenoid algae (Eu), Ascomycetes (Fa), Basidiomycetes (Fb), 
Deuteromycetes (Fd), Gymnosperm (Gp), Insects (In), Mammals (Ma), 
Oligochaeta (Ol), Agnatha (Pa), Chondrichthyes (Pc), Osteichthyes 
(Po), Protozoa (Pr), Reptiles (Re).

  \vspace{15 mm}
  
\textbf{Table \ref{dataset2}: Immunoglobulin-like fold proteins used as Data Set II.}

The first four columns contain the same type of information as in Table 
\ref{dataset1}.  Notice that the primary structures have been obtained 
from the \emph{PDB} data-bank in this case.  The 
full proteinsÕ names are listed in column 5.

 \vspace{15 mm}
 
 \textbf{Table \ref{clusters}: Clustering of the 20 natural AAs according to different criteria.}

The first two columns refer to the variable(s) upon which the 
clustering into $4$, $5$ or $10$ classes has been carried out by the 
\emph{k-means} algorithm.  In each case the value of the percent of 
the explained variability ($EV\%$) has been calculated as the 
following ratio : $EV\% = \frac{VarBetw}{VarBetw + VarWith}$, where 
$VarBetw$ and $VarWith$ are, respectively, the variability between the 
baricenters of the classes and the mean variability within each class.  
The last column provides an example of a "random clustering" of the 
$20$ AAs into the same number of classes.

\newpage


\begin{table}
\begin{footnotesize}
\caption{PCA on seven physico-chemical properties of the natural AAs. }\label{loadings}

\begin{center}
\def\V{\rule{0pt}{2.5ex}}
\begin{tabular} {lrrrrrrr}
\hline
\V & PC1 & PC2 & PC3 &  PC4 &  PC5 &  PC6 &  PC7 \\ [0.5ex]
\hline
EV\% \V & 50.04 &  34.73 &  7.43 &   5.29 &   1.90 &   0.47 &   0.14 \\ [0.5ex]
\hline
Hydrophobicity \V&  0.231 & 0.953 &   0.865 & -0.560 &  0.857 &  0.863 & -0.047 \\
Volume & -0.940 &  0.239 &  0.466 &  0.736 & -0.146 &  0.188 &  0.821 \\
Surface Area  &  -0.209 &  0.025 &  0.020 &  0.357 &  0.285 & -0.071 & -0.512 \\
Hydrophilicity &   0.052 &  0.067 & -0.023 & -0.017 &  0.362 & -0.423 &  0.229 \\
Bulkiness  &  0.023 & -0.142 & -0.172 &  0.064 &  0.180 &  0.192 &   0.096 \\
Refractivity  &   0.120 &  0.068 & -0.012 &  0.113 & -0.028 &  0.006 &  0.016 \\
Polarity  &   0.030 & -0.063 &  0.067 &  0.015 &  0.007 & -0.003 &  0.003 \\ [0.5ex]
\hline
\end{tabular}
\end{center}
\end{footnotesize}
\end{table}


\begin{table}
\begin{footnotesize}
\caption{Cytochromes of the \emph{c}  type used as Data Set I.}\label{dataset1}
\begin{center}
\def\V{\rule{0pt}{2.5ex}}
\begin{tabular}{rllr|rllr}
\hline
Id \V& Code & Species \bf{Fam.} & Length & Id & Code & Species \bf{Fam.} & 
Length \\ [1ex]
\hline

1  & ranca  & Rana Catesb. \bf{Am} &  104  &  36  & schpo  & Schizosac. Pombe \bf{Fa} &  108  \\
 2  & acene  & Acer Negun. \bf{Ap} &  112  &  37  & hanan  & Hansen. Anom. \bf{Fa} &  109  \\
 3  & fages  & Fagopyrum Escul. \bf{Ap} &  109  &  38  & issor  & Issatchen. Ori. \bf{Fa} &  109  \\
 4  & ricco  & Ricinus Comm. \bf{Ap} &  107  &  39  & neucr  & Neurosp. Cr. \bf{Fa} &  107  \\
 5  & braol  & Brassica Oler.  \bf{Ap} &  111  &  40  & torha  & Torulasp. Hans. \bf{Fa} &  109  \\
 6  & aruma  & Arum Macul.  \bf{Ap} &  109  &  41  & ustsp  & Ustilago Sphaer. \bf{Fb} &  107  \\
 7  & samni  & Sambucus Nig. \bf{Ap} &  111  &  42  & thela  & Thermomy. Lan. \bf{Fd} &  111  \\
 8  & cansa  & Cann. Sativa \bf{Ap} &  104  &  43  & ginbi  & Ginkgo Biloba \bf{Gp} &  107  \\
 9  & abuth  & Abutil. Theophr. \bf{Ap} &  108  &  44  & samcy  & Samia Cynthia \bf{In} &  107  \\
 10  & nigda  & Nigel. Damasc. \bf{Ap} &  101  &  45  & schgr  & Schistoc. Greg. \bf{In} &  107  \\
 11  & allpo  & Allium Porrum  \bf{Ap} &  105  &  46  & boepe  & Boettch. Per. \bf{In} &  107  \\
 12  & maize  & Zea Mays  \bf{Ap} &  109  &  47  & luccu  & Lucilia Cupr. \bf{In} &  107  \\
 13  & phaau  & Phaseolus Au. \bf{Ap} &  111  &  48  & apime  & Apis Mell. \bf{In} &  107  \\
 14  & troma  & Tropaeol. Majus \bf{Ap} &  109  &  49  & haeir  & Haematob. Irrit. \bf{In} &  107  \\
 15  & passa  & Pastin. Sativa \bf{Ap} &  107  &  50  & macma  & Macrobrac. Mal. \bf{In} &  104  \\
 16  & soltu  & Solanum Tuber. \bf{Ap} &  111  &  51  & manse  & Manduca Sexta \bf{In} &  107  \\
 17  & cucma  & Cucurb. Max. \bf{Ap} &  111  &  52  & canfa  & Canis Famil. \bf{Ma} &  104  \\
 18  & orysa  & Oryza Sativa \bf{Ap} &  111  &  53  & equas  & Equus Asinus \bf{Ma} &  104 \\
 19  & sesin  & Sesamum Indic. \bf{Ap} &  108  &  54  & horse  & Equus Caball. \bf{Ma} &  104  \\
 20  & gosba  & Gossypium Barbad. \bf{Ap}  &  108  &  55  & human  & Homo Sapiens \bf{Ma} &  104  \\
 21  & spiol  & Spinacia Oler. \bf{Ap} &  111  &  56  & minsc  & Miniopt. Schreib.  \bf{Ma} &  104  \\
 22  & helan  & Helianth. Ann. \bf{Ap} &  111  &  57  & macmu  & Macaca Mulat. \bf{Ma} &  104  \\
 23  & lyces  & Lycopersicon Escul. \bf{Ap} &  111  &  58  & atesp  & Ateles Sp. \bf{Ma} &  104  \\
 24  & wheat  & Triticum Aestiv.  \bf{Ap} &  112  &  59  & mirle  & Mirounga Leon. \bf{Ma} &  104  \\
 25  & astru  & Asterias Rub.  \bf{As} &  103  &  60  & eisfo  & Eisenia Foetida \bf{Ol} &  108  \\
 26  & chick  & Gallus Gallus \bf{Av} &  104  &  61  & enttr  & Entosphen. Trident. \bf{Pa} &  104  \\
 27  & anapl  & Anas Platyrhyn. \bf{Av} &  104  &  62  & squsu  & Squalus Sucklii  \bf{Pc} &  104  \\
 28  & drono  & Dromaius N.-Holl. \bf{Av} &  104  &  63  & cypca  & Cyprin. Carpio \bf{Po} &  94  \\
 29  & strca  & Struthio Camel. \bf{Av} &  104  &  64  & katpe  & Katsuwon. Pelamis \bf{Pr} &  103  \\
 30  & aptpa  & Aptenodytes Patag. \bf{Av} &  104  &  65  & crifa  & Crithidia Fasc.  \bf{Pr} &  113  \\
 31  & colli  & Columba Livia \bf{Av} &  104  &  66  & crion  & Crithidia Oncop. \bf{Pr} &  112  \\
 32  & helas  & Helix Aspersa \bf{Ga} &  98  &  67  & tetpy  & Tetrahymena Pyr. \bf{Pr} &  109  \\
 33  & chlre  & Chlamydom. Reinh. \bf{Ch} &  111  &  68  & croat  & Crotalus Atrox \bf{Re} &  104  \\
 34  & entin  & Enterom. Intest. \bf{Ch} &  100  &  69  & chese  & Chelydra Serp. \bf{Re} &  104  \\
 35  & euggr  & Euglena Gracil. \bf{Eu} &  102  &  &  &  &  \\ [1ex]
\hline
\end{tabular}
\end{center}
\end{footnotesize}

\end{table}
 


\begin{table}
\begin{footnotesize}
\caption{Immunoglobulin-like fold proteins used as Data Set II.}\label{dataset2}

\begin{center}
\def\V{\rule{0pt}{2.5ex}}
\begin{tabular}{rllrl}
\hline
No \V& PDB Id  & Source                  & Length & Protein's name \\ [1ex]
\hline
1  \V& 1ACX    & Actinomyces globisporus & 108 & Actinoxanthin   \\
2   & 1COB(A) & Bovine erythrocytes     & 151 & Superoxide dismutase \\ 
3   & 1CTM    & Turnip - Brassica rapa  & 250 & Cytochrome f  \\
4   & 1TEN    & Human                   & 90  & Tenascin     \\
5   & 3HHR(B) & Human                   & 197 & Human growth hormone   \\ 
6   & 3DPA    & Escherichia coli        & 218 & Pap D       \\
7   & 2RHE    & Human                   & 114 & Bence-Jones protein  \\
8   & 2MCG(1) & Human                   & 216 & Immunoglobulin lambda \\
9   & 1MCO(L) & Human                   & 216 & Immunoglobulin g1 \\
10  & 1FAI(L) & Mouse                   & 214 & Fab fragment \\
11  & 2FB4(H) & Human                   & 229 & Immunoglobulin fab \\
12  & 8FAB(B) & Human                   & 215 & Fab fragment \\
13  & 2FBJ(H) & Mouse                   & 220 & Ig A fab fragment \\
14  & 1CDB    & Mouse                   & 105 & T lymphocyte adesion glycoprotein \\
15  & 1TLK    & Turkey gizzard          & 103 & Telokin \\
16  & 1MCO(H) & Human                   & 428 & Immunoglobulin g1 \\
17  & 2IGE(A) & Human                   & 320 & Fc fragment (theoretical model) \\
18  & 1PFC    & Guinea pig serum        & 111 & Ig g1 P F c(prime) fragment \\
19  & 1CID    & Rat                     & 177 & T-cell surface glycoprotein Cd4 \\
20  & 3CD4    & Human                   & 178 & T-cell surface glycoprotein Cd4 \\
21  & 1DLH(A) & Human                   & 180 & Histocompatibility antigen Hla-dr1 \\
22  & 1DLH(B) & Human                   & 188 & Histocompatibility antigen Hla-dr1 \\
23  & 3HLA(A) & Human                   & 270 & Histocompatibility antigen Hla-a2 \\
\hline
\end{tabular}

\end{center}
\end{footnotesize}

\end{table}

\begin{table}
\begin{footnotesize}
\caption{Clustering of the 20 natural AAs according to different criteria.}\label{clusters}

\begin{center}
\def\V{\rule{0pt}{2.5ex}}
\begin{tabular} {rllll}
\hline
 & {\small Hydrophobicity} \V & {\small PC1 + PC2} & {\small Random} \\ [0.5ex]
\hline
\hline
4 Clusters \V& &  &  \\ [0.5ex]
\hline
1\V& A C G I L M F P S T W V  & C I L M P T V  & R  \\
2 & D E K  & R N D Q E H K  & N A S Y G H  \\
3 & N Q H Y  & A G S  & L M D F C W Q E  \\
4 & R  & F W Y  & I V K P T  \\[1.5ex]
EV : & 94\% & 84\% & --- \\ [0.5ex]
\hline
5 Clusters \V& &  &  \\ [0.5ex]
\hline
1\V& A C I L M F W V  & C I L M P T V  & M N  \\
2 & D E K  & N D Q E H  & D F Y P W E  \\
3 & N Q H  & A G S  & L V A K T H  \\
4 & R  & F W Y  & S R C  \\
5 & G P S T Y  & R K  & I G Q  \\[1.5ex]
EV : & 98\% & 90\% & --- \\ [0.5ex]
\hline
10 Clusters \V& &  &  \\ [0.5ex]
\hline
1\V& I L V  & I L M V  & K  \\
2 & D K  & N D  & M D P  \\
3 & N Q  & A S  & Q  \\
4 & R  & F Y  & L N G  \\
5 & A C W  & R K  & V T  \\
6 & P Y  & C P T  & E  \\
7 & H  & W  & A F  \\
8 & G S T  & G  & I Y H  \\
9 & M F  & Q H  & R C W  \\
10 & E  & E  & S  \\ [1.5ex]
EV : & 99.9\% & 98\% & --- \\ [0.5ex]
\hline
\end{tabular}
\end{center}
\end{footnotesize}
\end{table}


\end{document}